\documentclass[12pt]{article}
\usepackage{amssymb}
\title{\bf  An expanding $4D$ universe in a $5D$ Kaluza-Klein cosmology with
higher dimensional matter}
\author{F. Darabi\thanks{e-mail:
f.darabi@azaruniv.edu}\\{\small Department of Physics, Azarbaijan
University of Tarbiat Moallem, 53714-161, Tabriz, Iran .}\\
{\small Research Institute for Astronomy and Astrophysics of
Maragha, 55134-441, Maragha, Iran.} }
\begin{document}
\maketitle
\begin{abstract}
In the framework of Kaluza-Klein theory, we investigate a
$(4+1)$-dimensional universe consisting of a $(4+1)$ dimensional
Robertson-Walker type metric coupled with a $(4+1)$ dimensional
energy-momentum tensor. The matter part consists of an energy
density together with a pressure subject to $4D$ part of the
$(4+1)$ dimensional energy-momentum tensor. The dark part
consists of just a dark pressure $\bar{p}$, corresponding to the
extra-dimension endowed by a scalar field, with no element of
dark energy. It is shown that the reduced Einstein field
equations are free of $4D$ pressure and are just affected by an
effective pressure produced by the $4D$ energy density and dark
pressure. It is then proposed that the expansion of the universe
may be controlled by the equation of state in higher dimension
rather than four dimensions. This may account for the emergence
of unexpected current acceleration in the middle of matter
dominant era.
\end{abstract}
\newpage
\section{Introduction}

The recent distance measurements from the light-curves of several
hundred type Ia supernovae \cite{1,2} and independently from
observations of the cosmic microwave background (CMB) by the WMAP
satellite \cite{3} and other CMB experiments \cite{4,5} suggest
strongly that our universe is currently undergoing a period of
acceleration. This accelerating expansion is generally believed to
be driven by an energy source called dark energy. The question of
dark energy and the accelerating universe has been therefore the
focus of a large amount of activities in recent years. Dark energy
and the accelerating universe have been discussed extensively from
various point of views over the past few years
\cite{Quintessence,Phantom,K-essence}. In principle, a natural
candidate for dark energy could be a small positive cosmological
constant. One approach in this direction is to employ what is
known as modified gravity where an arbitrary function of the
Ricci scalar is added to the Einstein-Hilbert action. It has been
shown that such a modification may account for the late time
acceleration and the initial inflationary period in the evolution
of the universe \cite{modifidgravity2,modifidgravity}.
Alternative approaches have also been pursued, a few example of
which can be found in \cite{sahni,cardassian,chaplygin}. These
schemes aim to improve the quintessence approach overcoming the
problem of scalar field potential, generating a dynamical source
for dark energy as an intrinsic feature. The goal would be to
obtain a comprehensive model capable of linking the picture of
the early universe to the one observed today, that is, a model
derived from some effective theory of quantum gravity which,
through an inflationary period would result in today accelerated
Friedmann expansion driven by some $\Omega_{\Lambda}$-term.
However, the mechanism responsible for this acceleration is not
well understood and many authors introduce a mysterious cosmic
fluid, the so called dark energy, to explain this effect
\cite{Carroll}.

Since in a variety of inflationary models scalar fields have been
used in describing the transition from the quasi-exponential
expansion of the early universe to a power law expansion, it is
natural to try to understand the present acceleration of the
universe by constructing models where the matter responsible for
such behavior is also represented by a scalar field. Such models
are worked out, for example, in Ref \cite{6}.

Bellini {\it et al}, on the other hand, have published extensively
on the evolution of the universe from noncompact {\it vacuum}
Kaluza-Klein theory \cite{Bellini}. They used the essence of STM
(Space-Time-Matter) theory and developed a 5D mechanism to
explain, by a single scalar field, the evolution of the universe
including inflationary expansion and the present day observed
accelerated expansion. The STM theory is proposed by Wesson and
his collaborators, which is designed to explain the origin of
matter in terms of the geometry of the bulk space in which our
$4D$ world is embedded, for reviews see \cite{Wesson}. More
precisely, in STM theory, our world is a hypersurface embedded in
a five-dimensional Ricci-flat ($R_{AB}=0$) manifold where all the
matter in our world can be thought of as being manifestations of
the geometrical properties of the higher dimensional space
according to $G_{\alpha \beta}=8 \pi G T_{\alpha \beta}$,
provided an appropriate definition is made for the
energy-momentum tensor of matter in terms of the extra part of
the geometry. Physically, the picture behind this interpretation
is that curvature in $(4+1)$ space induces effective properties
of matter in $(3+1)$ space-time. The fact that such an embedding
can be done is supported by Campbell's theorem \cite{Compbell}
which states that any analytical solution of the Einstein field
equations in $N$ dimensions can be locally embedded in a
Ricci-flat manifold in $\left(N+1\right)$ dimensions. Since the
matter is induced from the extra dimension, this theory is also
called the induced matter theory. The sort of cosmologies
stemming from STM theory is studied in
\cite{LiuW,STM-cosmology,WLX}.

Another higher dimensional work has already been done with a
multi-dimensional {\it compact} Kaluza-Klein cosmological model in
which the scale factor of the compact space evolves as an inverse
power of the radius of the observable universe \cite{Mohamm}. The
Friedmann-Robertson-Walker equations of standard four-dimensional
cosmology were obtained where the pressure in the 4$D$ universe
was an effective pressure, expressed in terms of the components
of the higher dimensional energy-momentum tensor, capable of
being negative to explain the acceleration of our present
universe.

In this work, motivated by the work done in the compact model
\cite{Mohamm} and interested in its non-compact version, a 5$D$
{\it non-compact} Kaluza-Klein cosmological model is introduced
which is not Ricci flat, but is extended to couple with a higher
dimensional energy momentum tensor. In the present non-compact
model, it is shown that a dark pressure along the higher
dimensional sector together with the $4D$ energy density may
induce an effective pressure in four dimensional universe so that
the reduced field equations on $4D$ universe are free of $4D$
pressure and are just affected by the effective pressure. The
main point of this paper is to show the possibility that perhaps
$4D$ pressure does not directly control the dynamics of the
universe, rather the cosmological eras including inflation,
deceleration and current acceleration are just happened due to
either the evolution in equation of state along higher dimension
or an interplay between equations of state in $4D$ universe and
along higher dimension. Moreover, it is appealing to consider the
current acceleration of the universe as a result of a new phase
which is started recently along extra dimension. In this way, the
emergence of unexpected acceleration in the middle of matter
dominated era is easily justified because in this model the real
dynamics of the universe is controlled not by $4D$ physics but
through the higher dimensional physics. In other words, the
unexpected emergence of current acceleration may be related to a
higher dimensional effect which is hidden for $4D$ observers.

\section{The Model}

We start with the $5D$ line element
\begin{equation}
dS^2=g_{AB}dx^Adx^B, \label{0}
\end{equation}
in which $A$ and $B$ run over both the space-time coordinates
$\alpha, \beta$ and one non compact extra dimension indicated by
$4$. The space-time part of the metric $g_{\alpha \beta}=g_{\alpha
\beta}(x^{\alpha})$ is assumed to define the Robertson-Walker line
element
\begin{equation}
ds^2=dt^2-a^2(t)\left(\frac{dr^2}{(1-kr^2)}+r^2(d\theta^2+\sin^2\theta
d\phi^2 )\right), \label{1}
\end{equation}
where $k$ takes the values $+1, 0, -1$ according to a close, flat
or open universe, respectively. We also take the followings
$$
g_{4 \alpha}=0, \:\:\:\: g_{4 4}=\epsilon\Phi^2(x^{\alpha}),
$$
where $\epsilon^2=1$ and the signature of the higher dimensional
part of the metric is left general. This choice has been made
because any fully covariant $5D$ theory has five coordinate
degrees of freedom which can lead to considerable algebraic
simplification, without loss of generality.

The extra dimensional independence of the scalar field and the
$4D$ metric may pose an ambiguity between the compact or
non-compactness of the present model. In the compact theory the
cylindrical condition is imposed to account for this independence
and justify the non-observability of extra dimension. In
non-compact theory, however, one may account for this independence
within a different framework in which the non-observability of
extra dimensions is justified in a different way. Therefore, the
answer to the question of whether a higher dimensional model in
which all variables are independent of extra dimension is compact
or non-compact depends on the way by which the model justifies
the non-observability of extra dimensions. In the present model
we aim to follow the non-compact model where a reasonable
justification will be given for the non-observability of extra
dimension.

Unlike the noncompact vacuum Kaluza-Klein theory, we will assume
the fully covariant $5D$ non-vacuum Einstein equation
\begin{equation}
G_{AB}= 8 \pi G T_{AB}, \label{2}
\end{equation}
where $G_{AB}$ and $T_{AB}$ are the $5D$ Einstein tensor and
energy-momentum tensor, respectively. Note that the $5D$
gravitational constant has been fixed to be the same value as the
$4D$ one\footnote{In compact Kaluza-Klein theory one may define a
$5D$ gravitational constant $G^{(5)}$ which is reduced to $4D$
one as $G =\frac{G^{(5)}}{\int dy}$ where $\int dy$ is the volume
of extra compact dimension. In non-compact theory, however, we do
not require a $5D$ gravitational constant because there is no
finite volume of extra dimension, and assuming the gravitational
constant as $G$ will result in the correct $4D$ Einstein
equations.}.  In the following we use the geometric reduction
from 5$D$ to 4$D$ as appeared in \cite{Ponce}
\begin{equation}
\hat{R}_{\alpha \beta}={R}_{\alpha \beta}+\partial_4
\Gamma^4_{\alpha \beta}-\partial_{\beta} \Gamma^4_{\alpha 4}
+\Gamma^{\lambda}_{\alpha \beta}\Gamma^{4}_{\lambda
4}+\Gamma^{4}_{\alpha \beta}\Gamma^{D}_{4 D}-\Gamma^{4}_{\alpha
\lambda}\Gamma^{\lambda}_{\beta 4}-\Gamma^{D}_{\alpha
4}\Gamma^{4}_{\beta D},\label{5}
\end{equation}
where $\hat{}$ denotes the $4D$ part of the $5D$ quantities.
Evaluating the Christoffel symbols for the metric $g_{AB}$ gives
\begin{equation}
\hat{R}_{\alpha \beta}={R}_{\alpha
\beta}-\frac{\nabla_{\alpha}\nabla_{\beta}\Phi}{\Phi}.\label{6}
\end{equation}
In the same way we obtain
\begin{equation}
{R}_{4 4}=-\epsilon\Phi \Box \Phi. \label{8}
\end{equation}
We now construct the space-time components of the Einstein tensor
$$
G_{AB}=R_{AB}-\frac{1}{2}g_{AB}R_{(5)}.
$$
In so doing, we first obtain the $5D$ Ricci scalar $R_{(5)}$ as
$$
R_{(5)}=g^{AB}R_{AB}= \hat{g}^{\alpha \beta}
\hat{R}_{\alpha\beta}+ g^{44}R_{44}= g^{\alpha \beta}(R_{\alpha
\beta}-\frac{\nabla_{\alpha}\nabla_{\beta}\Phi}{\Phi})+\frac{\epsilon}{\Phi^2}(-\epsilon\Phi
\Box \Phi)
$$
\begin{equation}
=R-\frac{2}{\Phi}\Box\Phi,\label{9}
\end{equation}
where the $\alpha 4$ terms vanish and $R$ is the $4D$ Ricci
scalar. The space-time components of the Einstein tensor is
written $\hat{G}_{\alpha \beta}=\hat{R}_{\alpha
\beta}-\frac{1}{2}\hat{g}_{\alpha \beta}R_{(5)}$. Substituting
$\hat{R}_{\alpha \beta}$ and $R_{(5)}$ into the space-time
components of the Einstein tensor gives
\begin{equation}
\hat{G}_{\alpha \beta}={G}_{\alpha \beta}+\frac{1}{\Phi}(g_{\alpha
\beta} \Box \Phi- \nabla_{\alpha}\nabla_{\beta}\Phi). \label{10}
\end{equation}
In the same way, the 4-4 component is written ${G}_{4 4 }={R}_{4 4
}-\frac{1}{2}g_{4 4}R_{(5)}$, and substituting ${R}_{4 4}$,
$R_{(5)}$ into this component of the Einstein tensor gives
\begin{equation}
G_{4 4}=-\frac{1}{2}\epsilon R\Phi^2. \label{11}
\end{equation}
We now consider the $5D$ energy-momentum tensor without specifying
its nature or origin\footnote{Since this model is supposed to
describe the radiation and matter dominated eras, it seems that
this fifth dimensional pressure component would be a {\it dark}
property of conventional matter, including standard model fields
(see refs in \cite{Randj} for some derivations of the matter
contribution in Kaluza-Klein cosmology).}. The form of
energy-momentum tensor is dictated by Einstein's equations and by
the symmetries of the metric (\ref{1}). Therefore, we may assume
a perfect fluid with nonvanishing elements
\begin{equation}
{T}_{\alpha \beta}=(\rho+p){u}_{\alpha} {u}_{\beta}-p{g}_{\alpha
\beta}, \label{12}
\end{equation}
\begin{equation}
{T}_{44}=-\bar{p}g_{44}= -\epsilon\bar{p}\Phi^2, \label{14}
\end{equation}
where $\rho$ and $p$ are the conventional density and pressure of
perfect fluid in the $4D$ standard cosmology and $\bar{p}$ acts
as a pressure living along the higher dimensional sector. Notice
that the perfect fluid is isotropic on the 3$D$ geometry and
anisotropic regarding the 5$^{th}$ dimension\footnote{The same
choice has been made in \cite{Mohamm} with the components of the
higher dimensional energy-momentum tensor as $T_{ij}=diag[-\rho,
p, p, p, p_d, ..., p_d]$ where $\rho, p$ and $p_d$ are the
density, pressure on the 3$D$ geometry and pressure along the
extra dimensions, respectively. }. The field equations (\ref{2})
are to be viewed as {\it constraints} on the simultaneous
geometric and physical choices of $G_{AB}$ and $T_{AB}$
components, respectively.

Substituting the energy-momentum components (\ref{12}), (\ref{14})
in front of the $4D$ and extra dimensional part of Einstein
tensors (\ref{10}) and (\ref{11}), respectively, we obtain the
field equations\footnote{The $\alpha 4$ components of Einstein
equation (\ref{2}) result in
$$
{R}_{\alpha 4}=0,
$$
which is an identity with no useful information.}
\begin{equation}
G_{\alpha \beta}=8 \pi G [(\rho+p)u_{\alpha} u_{\beta}-pg_{\alpha
\beta}]+\frac{1}{\Phi}\left[\nabla_{\alpha}\nabla_{\beta}\Phi-\Box
\Phi g_{\alpha \beta}\right], \label{15}
\end{equation}
and
\begin{equation}
R=16 \pi G \bar{p}.\label{16}
\end{equation}
By evaluating the $g^{\alpha \beta}$ trace of Eq.(\ref{15}) and
combining with Eq.(\ref{16}) we obtain
\begin{equation}
\Box\Phi=\frac{1}{3}(8\pi G(\rho-3p)+16 \pi G \bar{p})\Phi
.\label{18}
\end{equation}
This equation infers the following scalar field potential
\begin{equation}
V(\Phi)=-\frac{1}{6}(8\pi G(\rho-3p)+16 \pi G \bar{p})\Phi^2,
\end{equation}
whose minimum occurs at $\Phi=0$, for which the equations
(\ref{15}) reduce to describe a usual $4D$ FRW universe filled
with ordinary matter $\rho$ and $p$. In other words, our
conventional $4D$ universe corresponds to the vacuum state of the
scalar field $\Phi$. From Eq.(\ref{18}), one may infer the
following replacements for a nonvanishing $\Phi$
\begin{equation}
\frac{1}{\Phi}\Box \Phi = \frac{1}{3}(8\pi G(\rho-3p)+16 \pi G
\bar{p}),\label{19}
\end{equation}
\begin{equation}
\frac{1}{\Phi}\nabla_{\alpha}\nabla_{\beta}\Phi = \frac{1}{3}(8\pi
G(\rho-3p)+16 \pi G \bar{p})u_{\alpha}u_{\beta}.\label{20}
\end{equation}
Putting the above replacements into Eq.(\ref{15}) leads to
\begin{equation}
G_{\alpha \beta}=8 \pi G [(\rho+\tilde{p})u_{\alpha}
u_{\beta}-\tilde{p}g_{\alpha \beta}], \label{22}
\end{equation}
where
\begin{equation}
\tilde{p}=\frac{1}{3}(\rho+2\bar{p}).\label{23}
\end{equation}
This energy-momentum tensor effectively describes a perfect fluid
with density $\rho$ and pressure $\tilde{p}$. The four dimensional
field equations lead to Friedmann equation
\begin{equation}
3\frac{\dot{a}^2+k}{a^2}=8 \pi G \rho, \label{24}
\end{equation}
and
\begin{equation}
\frac{2a\ddot{a}+\dot{a}^2+k}{a^2}=-8 \pi G \tilde{p}. \label{25}
\end{equation}
Differentiating (\ref{24}) and combining with (\ref{25}) we obtain
the conservation equation
\begin{equation}
\frac{d}{dt}(\rho a^3)+\tilde{p}\frac{d}{dt}(a^3)=0. \label{26}
\end{equation}
The equations (\ref{24}) and (\ref{25}) can be used to derive the
acceleration equation
\begin{equation}
\frac{\ddot{a}}{a}=-\frac{4 \pi G}{3}(\rho+3\tilde{p})=-\frac{8
\pi G}{3}(\rho+\bar{p}). \label{27}
\end{equation}
The acceleration or deceleration of the universe depends on the
negative or positive values of the quantity $(\rho+\bar{p})$.
\\From extra dimensional equation (\ref{16}) ( or $4$-dimensional
Eqs.(\ref{23}), (\ref{24}) and (\ref{25}) ) we obtain
\begin{equation}
-\frac{6(k+\dot{a}^2+\ddot{a}a)}{a^2}=16 \pi G \bar{p}.\label{28}
\end{equation}
Using power law behaviors for the scale factor and dark pressure
as $a(t)=a_0t^{\alpha}$ and $\bar{p}(t)=\bar{p}_0t^{\beta}$ in the
above equation, provided $k=0$ in agreement with observational
constraints, we obtain $\beta=-2$.

Based on homogeneity and isotropy of the 4$D$ universe we may
assume the scalar field to be just a function of time, then the
scalar field equation (\ref{18}) reads as the following form
\begin{equation}
\ddot{\Phi}+3\frac{\dot{a}}{a}\dot{\phi}-\frac{8\pi
G}{3}((\rho-3p)+2 \bar{p})\Phi=0 .\label{29}
\end{equation}
Assuming $\Phi(t)=\Phi_0t^{\gamma}$ and $\rho(t)=\rho_0t^{\delta}
\: (\rho_0>0)$ together with the equations of state for matter
pressure $p=\omega \rho$ and dark pressure $\bar{p}=\Omega \rho$
we continue to calculate the required parameters for inflation,
deceleration and then acceleration of the universe\footnote{As we
discussed earlier, the fifth dimensional pressure component could
be a {\it dark} property of conventional matter $\rho$, through a
dark parameter $\Omega$, according to $\bar{p}=\Omega \rho$.}. In
doing so, we rewrite the acceleration equation (\ref{27}), scalar
field equation (\ref{29}) and conservation equation (\ref{26}),
respectively, in which the above assumptions are included as
\begin{equation}
\alpha(\alpha -1)+\frac{8 \pi G}{3}\rho_0(1+\Omega)=0, \label{30}
\end{equation}
\begin{equation}
\gamma(\gamma -1)+3\alpha \gamma-\frac{8\pi
G}{3}\rho_0((1-3\omega)+2 \Omega)=0 ,\label{31}
\end{equation}
\begin{equation}
2\rho_0[(2+\Omega)\alpha-1]=0, \label{32}
\end{equation}
where $\delta=-2$ has been used due to the consistency with the
power law behavior $t^{3\alpha -3}$ in the conservation equation.
The demand for acceleration $\ddot{a}>0$ through Eq.(\ref{27})
with the assumptions $\rho(t)=\rho_0t^{\delta}$ and
$\bar{p}=\Omega \rho$, requires $\rho_0(1+\Omega)<0$ or
$\Omega<-1$ which accounts for a negative dark pressure. This
negative domain of $\Omega$ leads through the conservation
equation (\ref{26}) to $\alpha>1$ which indicates an accelerating
universe as expected. On the other hand, using Friedmann equation
we obtain $\alpha=\frac{1}{2+\Omega}$ which together with the
condition $\alpha>1$ requires that  $-2<\Omega<-1$. Now, one may
recognize two options as follows.

The first option is to attribute an intrinsic evolution to the
parameter $\Omega$ along the higher dimension so that it can
produce the $4D$ expansion evolution in agreement with standard
model including early inflation and subsequent deceleration, and
also current acceleration of the universe. Ignoring the
phenomenology of the evolution of the parameter $\Omega$, we may
require
\begin{equation} \label{36}
\left\{ \begin{array}{ll} \Omega \gtrsim -2 \:\:\:\:\:\:\:\:
{for} \:\:\: \mbox{inflation}
\\
{\Omega}>-1 \:\:\:\:\:\:\:\: {for} \:\:\: \mbox{deceleration}
\\
{\Omega}\lesssim-1 \:\:\:\:\:\:\:\: {for} \:\:\:
\mbox{acceleration}.
\end{array}
\right.
\end{equation}
The first case corresponds to highly accelerated universe due to
a large $\alpha>>> 1$. This can be relevant for the inflationary
era if one equate the power law with exponential behavior. The
second case corresponds to a deceleration $\alpha< 1$, and the
third case represents an small acceleration $\alpha \gtrsim 1$.
In this option, there is no specific relation between the
physical phase along extra dimension, namely $\Omega$, and the
ones defined in $4D$ universe by $\omega$.

The second option is to assume a typical relation between the
parameters $\Omega$ and $\omega$ as $\Omega=f(\omega)$ so that
\begin{equation} \label{38}
\left\{ \begin{array}{ll} \Omega \gtrsim -2 \:\:\:\:\:\:\:\:
{for} \:\:\: \omega ={-1}
\\
{\Omega}>-1 \:\:\:\:\:\:\:\: {for} \:\:\: \omega=\frac{1}{3}
\\
{\Omega}\lesssim-1 \:\:\:\:\:\:\:\: {for} \:\:\: \omega=0.
\end{array}
\right.
\end{equation}
The case $\omega = -1$  corresponds to the early universe and
shows a very high acceleration due to $\alpha>>> 1$. The case
$\omega=\frac{1}{3}$ corresponds to the radiation dominant era
and shows a deceleration $\alpha< 1$. Finally, the case
$\omega=0$ corresponds to the matter dominant era and shows an
small acceleration $\alpha \gtrsim 1$ in agreement with
observations.

\section*{Conclusion}

A $(4+1)$-dimensional universe consisting of a $(4+1)$ dimensional
metric of Robertson-Walker type coupled with a $(4+1)$ dimensional
energy-momentum tensor in the framework of noncompact Kaluza-Klein
theory is studied. In the matter part, there is energy density
$\rho$ together with pressure $p$ subject to $4D$ part of the
$(4+1)$ dimensional energy-momentum tensor, and a dark pressure
$\bar{p}$ corresponding to the extra-dimensional part endowed by a
scalar field. A particular (anisotropic) equation of state in $5D$
is used for the purpose of realizing the $4D$ expansion in
agreement with observations. This is done by introducing two
parameters $\omega$ and $\Omega$ which may be either independent
or related as $\Omega=f(\omega)$. The physics of $\omega$ is well
known but that of the parameter $\Omega$ needs more careful
investigation based on effective higher dimensional theories like
string theory or Brane theory. The reduced $4D$ and
extra-dimensional components of $5D$ Einstein equations together
with different equations of state for pressure $p$ and dark
pressure $\bar{p}$ may lead to a $4D$ universe which represents
early inflation, subsequent deceleration and current
acceleration. In other words, all eras of cosmic expansion may be
explained by a single simple mechanism.

The important point of the present model is that the reduced
Einstein field equations are free of $4D$ pressure and are just
affected by an effective pressure produced by the $4D$ energy
density and dark pressure along the extra dimension. This provides
an opportunity to consider the expansion of the universe as a
higher dimensional effect and so justify the {\it unexpected
current acceleration in the middle of matter dominant era}, along
this line of thought. Moreover, there is no longer ``coincidence
problem'' within this model. This is because, in the present
model there is no element of ``dark energy'' at all and we have
just one energy density $\rho$ associated with ordinary matter.
So, there is no notion of coincidental domination of dark energy
over matter densities to trigger the acceleration at the present
status of the universe. In fact, a dark pressure with different
negative values along the $5^{th}$ dimension by itself may produce
expanding universe including inflation, deceleration and
acceleration without involving with the coincidence problem.
These stages of the 4$D$ universe may occur as well because of
negative, positive and zero values of the four dimensional
pressure, respectively, which leads to a competition between
energy density $\rho$ and dark pressure $\bar{p}$ in the
acceleration equation (\ref{27}). For the same reason that there
is no element of dark energy in this model, the apparent {\it
phantom like} equation of state for dark pressure $\Omega<-1$ is
free of serious problems like {\it unbounded from below dark
energy} or {\it vacuum instability} \cite{Cline}.

The above results are independent of the signature $\epsilon$ by
which the higher dimension takes part in the 5D metric. Moreover,
the role played by the scalar field along the $5^{th}$ coordinate
in the $5D$ metric is very impressed by the role of scale factor
over the $4D$ universe. At early universe during the inflationary
era the scalar field is highly suppressed and the $5^{th}$
coordinate is basically ignored in $5D$ line element. At
radiation dominant era the scalar field is much less suppressed
and the $5^{th}$ coordinate becomes considerable in $5D$ line
element. Finally, at matter dominant era the scalar field and its
possible fluctuations starts to be super-suppressed and the
observable effect of $5^{th}$ coordinate becomes vanishing in
$5D$ line element at $t\simeq 10^{17}{Sec}$, leaving an effective
$4D$ universe in agreement with observations.

A clear similarity is seen between the results of our 5$D$ {\it
non compact} model and that of multi-dimensional {\it compact} one
\cite{Mohamm}. Both of these models predict an effective 4$D$
pressure, expressed in terms of the components of the higher
dimensional energy-momentum tensor, capable of being negative to
explain the acceleration of our present universe. Moreover, both
higher dimensional metrics dynamically evolves towards an
effective four-dimensional one.

\section*{Acknowledgment}

This work has been financially supported by the Research Institute
for Astronomy and Astrophysics of Maragha (RIAAM).

\end{document}